\input{aipcheck}

\documentclass[
    ,final            
  ]
  {aipproc}

\layoutstyle{8x11single}

\begin{document}

\title{Partition Function of Interacting Calorons Ensemble}


\author{S.~Deldar }{
  address={Department of Physics, University of Tehran, P.O. Box 14395/547, Tehran 1439955961,
Iran}
}
\author{M.~Kiamari}{
altaddress={Department of Physics, University of Tehran, P.O. Box 14395/547, Tehran 1439955961,
Iran}
}

\begin{abstract}

We present a method for computing the partition function of  a caloron ensemble taking into account  the interaction of calorons. We focus on caloron-Dirac string interaction and show 
that the metric that Diakonov and Petrov offered works well in the limit where this interaction occurs. We suggest computing the correlation function of two polyakov loops by applying 
Ewald's method.

\end{abstract}

\maketitle

\section{INTRODUCTION}

KvBLL instantons or calorons with non-trivial holonomy can be taken as the degrees of freedom in Yang-Mills vacuum in any temperature. The non-interacting ensemble of calorons has 
been studied by Diakonov and Petrov \cite{Diakonov2009}. They obtained the heavy quark potential by calculating the correlation function of two Polyakov loops:
\begin{equation}
{e^{ - \beta \Delta {F_{q\bar q}}}} = \left\langle {TrL\left(\textbf{r}  \right)Tr{L^\dag }(\textbf{r})} \right\rangle.
\label{potential}
\end{equation}
To compute the correlation function, the partition function of caloron ensemble has to be calculated. Therefore, one has to find the contribution of a caloron in the partition function.
Diakonov and Gromov \cite{Diakonov2005} computed this contribution using the metric of the caloron moduli space in terms of N different SU(N) BPS monopoles
\[Z = {e^{ - {S_{cl}}}}\int {J\prod\limits_p {\frac{{d{Y_P}}}{{\sqrt {2\pi } g}}} } \int {D{a_\mu }D\chi D\bar \chi } \exp \left( { - \frac{1}{{2{g^2}}}\int {{d^4}xa_\mu ^aW_{\mu \nu }^{ab}a_\nu ^b}  - \int {{d^4}x{{\bar \chi }^a}{{\left( {{D^2}} \right)}^{ab}}{\chi ^b}} } \right).\]
Where $ Y_P $ represents the  caloron collective coordinates and the Jacobian J is the determinant of the moduli space metric tensor. Diakonov and \textit{et al}.\cite{Diakonov2004} computed the full partition function for KvBLL instanton in SU(2) gauge theory:
\begin{equation}
\begin{array}{l}
Z = \mathop \smallint \nolimits {d^3}{z_1}{d^3}{z_2}d{\cal O}{T^6}C{\left( {\frac{{8{\pi ^2}}}{{{g^2}}}} \right)^4} \exp \left[ { - VP\left( \nu  \right) - 2\pi {r_{12}}P''\left( \nu  \right)} \right]\\
 \times {\left( {\frac{{\Lambda {e^{{\gamma _E}}}}}{{4\pi T}}} \right)^{\frac{{22}}{3}}}{\left( {\frac{1}{{T{r_{12}}}}} \right)^{\frac{5}{3}}} \left( {2\pi  + \frac{{\nu \bar \nu }}{T}{r_{12}}} \right){\left( {\nu {r_{12}} + 1} \right)^{\frac{{4\nu }}{{3\pi T}} - 1}}{\left( {\bar \nu {r_{12}} + 1} \right)^{\frac{{4\bar \nu }}{{3\pi T}} - 1}} 
\end{array}
\label{partition}
\end{equation}
where C  is a constant and
\[P(\nu ) = \frac{1}{{12{\nu ^2}T}}{\nu ^2}{{\bar \nu }^2},P''(\nu ) = \frac{1}{{{\pi ^2}T}}[\pi T(1 - \frac{1}{{\sqrt 3 }}) - \nu ][\bar \nu  - \pi T(1 - \frac{1}{{\sqrt 3 }})].\]
Using the partition function Z, for computing the correlation function of Polyakov loops in eq. \ref{potential}, one can obtain the potential between a quark-anti quark pair.
\\
In this research we try to add the contribution  of the interacting calorons to the calculations. As a result of this change, the metric of the interacting calorons moduli space should 
be modified. However, one may keep the same moduli space with the same metric as the non interacting calorons, but introducing a potential on the moduli space which corresponds to 
the interaction between calorons. The main task of this research is to give some ideas about computing the partition function of the interacting calorons and finally calculating  the potential 
between static quarks. However, one can compute correlation function of the Polyakov loops without using metric by applying Ewald's method.
\\
In the next two sections we use the sum ansatz idea and the caloron-Dirac string interaction to suggest how one may calculate the partition function. In section 4, we discuss about the
caloron-anticaloron ansatz and in section 5, we argue about the interval where the metric of Diakonov and Petrov is positive definite and we explain that our calculations in section 3 is 
done in this interval. In section 6, we apply Ewald's method to caloron ensemble for large caloron radius limit.

\section{Sum Ansatz}
\label{2}

Even though, two calorons do not interact at large distances in algebraic gauge but they may interact when they get close to each other.
\\
Consider two calorons with non-trivial holonomies. Because $ A_4 $'s of these two calorons do not vanish even at spatial infinity, calorons correlate with each other at infinite separation. 
Therefore one should not simply add two single caloron gauge fields. It is possible to choose algebraic gauge, in which all components of potential of a caloron vanish at spatial infinity, and
then add the potentials of calorons:
\begin{equation}
{{A_\mu } = A_\mu ^{(1)} + A_\mu ^{(2)}}.
\label{amu}
\end{equation}
We recall that to compute the partition function of sum ansatz, $ A_\mu $'s in eq. \ref{amu} should be transformed back to the periodic gauge. This gauge transformation is dependent to 
the holonomy, therefore two calorons should have identical holonomies. The field strength of this ansatz can be written as
\[{H_{\mu \nu }}\left( {{A^{\left( 1 \right)}} + {A^{\left( 2 \right)}}} \right) = {F_{\mu \nu }}\left( {{A^{\left( 1 \right)}}} \right) + {F_{\mu \nu }}\left( {{A^{\left( 2 \right)}}} \right) + {F_{\mu \nu }}\left( {{A^{\left( 1 \right)}},{A^{\left( 2 \right)}}} \right),\]
where the additional term (the third one on the right) is due to the non-linear nature of the theory. Calculating the action, we have an extra term in addition to the action of individual 
calorons that can be interpreted as the effective classical potential corresponding to the interaction between two calorons 
\begin{equation}
{\exp \left( -{S_{int}} \right) \equiv exp\left( { - \frac{1}{{4{g^2}}}\mathop \smallint \nolimits {d^4}x\left( {H_{\mu \nu }^2 - \mathop \sum \limits_{i = 1}^2 F_{\mu \nu }^2\left( A^{(i)} \right)} \right)} \right)}.
\label{interaction}
\end{equation}
Now we should write the partition function of two interacting calorons. We should note that in computing the partition function we change the dominant path to combination of two calorons in periodic gauge, so we must consider the quantum fluctuation about this new path and compute the metric and quantum potential. But as the first order of approximation we can use the metric of one individual caloron and write the partition function as:
\begin{equation}
{Z_{2c}} = \frac{1}{{2!}} \mathop\prod \limits_{i = 1}^2 {Z_i}{e^{ - {S_{int}}}},
\label{z2c}
\end{equation}
where $ Z_i $ has been derived in eq. \ref{partition} .
\\
Now we study the ensemble of K calorons. To simplify the calculations we restrict ourselves to a system which consists of many pairs of calorons. The calorons in each pair interact with 
each other but each pair does not interact with  the others. The grand canonical partition function of these K pairs can be written as:
\[Z = \mathop \sum \limits_K {f^K}{Z_K} = \mathop \sum \limits_K \frac{{{{(f{Z_{2c}})}^K}}}{{K!}}.\]
Where f is the fugacity and should be calculated, and $Z_{2c}$ can be obtained from eq. \ref{z2c}. With this partition function, one can calculate the correlation function of Polyakov loops and then compute the heavy quarks potential.
\\
In the next section, we argue about another alternative to include the interaction between calorons in the partition function.

\section{ Caloron-Dirac String Interaction}
\label{3}
The next idea for computing the partition function of interacting calorons is using the caloron-Dirac string interaction which has been studied by Gerhold \textit{et al}. \cite{Gerhold2007}. The interaction between the caloron and the Dirac string had not been taken into account in the non interacting caloron partition function and therefore this is a new effect in computing the partition function of interacting calorons where two calorons are located on top of each other. It means that two calorons interact with each other by their Dirac strings.
\\
In the large caloron radius limit, $ \rho \gg \beta $ , where $ \rho $ is the radius of the caloron,  the vector potential of the caloron in algebraic gauge is dominantly Abelian and governed by the 
third component in color space. For convenience, the monopole axis is taken along the third axis.
\[\left| {A_\mu ^{(1,2)}\left( {\tilde x} \right)} \right| \ll A_{1,2}^{(3)}\left( {\tilde x} \right),\left| {A_{3,4}^{(3)}\left( {\tilde x} \right)} \right| \ll A_{1,2}^{(3)}\left( {\tilde x} \right)\]
This dominant component is called a Dirac string.
\\
Since the interaction due to the Dirac string is not negligible, when an object (monopole or caloron) is placed on the top of the Dirac string of a caloron, the ordinary sum ansatz can not be 
used for the superposition of caloron and this object. Gerhold \textit{et al}.\cite{Gerhold2007} derived the superposition by removing the Dirac string in the algebraic gauge doing a proper gauge 
transformation (G), then they added the gauge fields of this caloronand that object and finally applied the inverse gauge transformation to have the original Dirac string back. The result is: 
\begin{equation}
{A_\mu ^{final}\left( x \right) = {A_\mu }^{Cal}\left( x \right) + {e^{ - iG(x){\tau _3}}}{A_\mu }^{Obj}\left( x \right){e^{iG(x){\tau _3}}}},
\label{rotation}
\end{equation}
Therefore the effect of the Dirac string is included by adding an unchanged caloron to a gauge rotated object.
\\
Now consider two interwined calorons such that  one of the monopoles of a caloron is located on the top of the Dirac string of the other caloron and vice versa. In this case one should 
rotate only the monopole of each caloron which touches the Dirac string of the other caloron  as explained in eq. \ref{rotation}. These rotations do not affect the Dirac strings, because monopoles rotate around the 
monopole axis which is along the third axis. We should recall that  the Dirac strings are also defined along the third axis. Applying these rotations, one can repeat the approach of the 
previous section and obtain the field strength of 
$ A_\mu ^{final} $ and use eq. \ref{interaction} to compute the interaction term.
Again, we  restrict ourselves to a system which consists of many pairs of calorons. The calorons in each pair interact with each other but each pair does not interact with the others. We 
can compute the partition function of K non-interacting pairs of calorons as in previous section.

\section{Caloron-Anticaloron Ansatz}
\label{4}
The  sum ansatz idea can be applied to a caloron and anticaloron system. Although the combination of solutions of caloron and anticaloron is not the solution of equation of motion but
both of these objects minimize the action and may have identical integral measure and partition functions. They may be saddle points of the partition function. In 1984, Diakonov and 
Petrov \cite{Diakonov1984} argued  about the instanton-anti instanton solutions and wrote:"Thus, from physical considerations it is clear that the main contribution to the partition function should be 
given not by exact classical solutions, but by approximate solutions of the instanton-anti instanton type. This should not surprise us since the approximate solutions may have a larger 
entropy (= statistical weight) than the exact solutions which correspond to the local minima of the action. Anyhow, one is interested in maximizing the partition function of a theory, and 
instanton-anti instanton configurations may well give a larger contribution to it than those of multi-instantons." Therefore, there is a possibility of constructing the partition function from 
caloron anticaloron pairs instead of calorons or anticalorons only. However, we recall that the caloron and anticaloron should have the same holonomy if one wants to do an appropriate gauge 
transformation, from algebraic gauge to the periodic gauge,  on the combination.  .
\\
Since the function G which is applied to remove the Dirac string of anticaloron is the same as the caloron's 
\cite{Gerhold2007}, we can apply the caloron-Dirac string interaction to the caloron-anticaloron ansatz. Thus, we can consider an ensemble with equal numbers of calorons and anticalorons 
in which one caloron interacts with only one anticaloron (either interaction  due to the ordinary sum ansatz or caloron-Dirac string interaction). Then, we again assume  that  these pairs do not 
interact and we use the same procedure as the previous section to calculate the partition function. 

\section{The Metric of Moduli Space}
 \label{5}
In 2009, Bruckmann \textit{et al}. \cite{Bruckmann2009} showed that the metric of moduli space of dyons which Diakonov and Petrov \cite{Diakonov2007} offered, does not have the essential 
requirement of positive definiteness, throughout the whole configuration space. They argue that this metric is positive definite only for the low dyon density.
\\
We recall the approximate metric of the same kind dyons in SU(2) is \cite{Diakonov2007}:
\[G = \left( {\begin{array}{*{20}{c}}
{2\pi  - {\raise0.7ex\hbox{$2$} \!\mathord{\left/
 {\vphantom {2 d}}\right.\kern-\nulldelimiterspace}
\!\lower0.7ex\hbox{$d$}}}&{ + {\raise0.7ex\hbox{$2$} \!\mathord{\left/
 {\vphantom {2 d}}\right.\kern-\nulldelimiterspace}
\!\lower0.7ex\hbox{$d$}}}\\
{ + {\raise0.7ex\hbox{$2$} \!\mathord{\left/
 {\vphantom {2 d}}\right.\kern-\nulldelimiterspace}
\!\lower0.7ex\hbox{$d$}}}&{2\pi  - {\raise0.7ex\hbox{$2$} \!\mathord{\left/
 {\vphantom {2 d}}\right.\kern-\nulldelimiterspace}
\!\lower0.7ex\hbox{$d$}}}
\end{array}} \right),\]
The eigenvalues of this metric are:
\[{\lambda _1} = 2\pi ,{\lambda _2} = 2\pi  - \frac{4}{d},\]
where d is the dyon separation. Det (G), which represents the weight factor, is positive for $ d > \frac{2}{\pi } \equiv \frac{2}{{\pi T}}  $ , where T shows 
the temperature. Therefore for the dyon separation smaller than this critical distance, where two dyons overlap, this approximate metric does not work and one should refine it. 
(The metric of different kind dyons is positive definite for all dyon separations, because the minus sign in the above formula changes to plus sign and therefore the eigenvalues of the metric are positive for all d).
But we recall that for the  caloron-Dirac string interaction, we take the caloron in the large raduis limit, $ \rho \gg \beta $ , where the vector potential of the caloron is dominantly abelian. 
Rewriting the limit of positive definiteness,
\[d = \frac{{\pi {\rho ^2}}}{\beta } > 2\frac{\beta }{\pi },\]
yielding $ \frac{\rho }{\beta } > \frac{{\sqrt 2 }}{{{\pi ^2}}} \approx 0.45 $ . Therefore in the limit where we need to consider the Dirac string, the metric which Diakonov and Petrov offered is positive definite and works well.
We shoule notice that for calculating the partition function from sum ansatz, one should either consider the calorons in the large raduis limit or modify the metric.

\section{Ewald's method}
 \label{6}
To deal with the problem of positive definiteness, Bruckmann \textit{et al}. \cite{Bruckmann2012} used the Ewald's method to compute the correlation function of two polyakov loops in 
eq.\ref{potential} for non-interacting ensemble. The vector potential for 2K dyons ensemble in the abelian limit is
\[{A_\mu }(\textbf{r}) = \left( {{\delta _{\mu 0}}2\pi \omega T + \frac{1}{2}\sum\limits_{i = 1}^K {\sum\limits_{m = 1}^2 {{a_\mu }(\textbf{r} -\textbf{r} _i^m;{q_m})} } } \right){\sigma _3},\]
where $ {a_\mu }(\textbf{r};q) =  - q\bar \eta _{\mu \nu }^3{\partial _\nu }ln(r - z). $
\\
Then, one  rewrites the polyakov loop 
\[\begin{array}{l}
L\left(\textbf{r} \right) \equiv \frac{1}{2}Tr\left( {\exp \left( {i\mathop \smallint \limits_0^{1/T} d{x_0}{A_0}\left( {{x_0},\textbf{r}} \right)} \right)} \right) \to L\left(\textbf{r}  \right) = \cos \left( {2\pi \omega  + \frac{1}{T}\Phi \left(\textbf{r} \right)} \right)\\
 \to {\left. {L\left(\textbf{r}\right)} \right|_{\omega  = {\raise0.7ex\hbox{$1$} \!\mathord{\left/
 {\vphantom {1 4}}\right.\kern-\nulldelimiterspace}
\!\lower0.7ex\hbox{$4$}}}} = \ - sin \left( {\frac{1}{2T}\Phi \left(\textbf{r} \right)} \right)
\end{array},\]
where $ \Phi (\textbf{r}) \equiv \sum\limits_{i = 1}^K {\sum\limits_{m = 1}^2 {\frac{{q_m}}{{|\textbf{r} -\textbf{r} _i^m|}}} } $ is a long-rang function. For maximal non-trivial holonomy, both 
dyons of the SU(2) gauge group have the same action. Therefore
\[\left\langle O \right\rangle  = \frac{{\mathop \smallint \nolimits \mathop \prod \limits_{i = 1}^K d\textbf{r}_i^1d\textbf{r}_i^2O\left( {\left\{ {\textbf{r}_i^1,\textbf{r}_i^2} \right\}} \right)}}{{\mathop \smallint \nolimits \mathop \prod \limits_{i = 1}^K d\textbf{r}_i^1d\textbf{r}_i^2}} = \frac{{\mathop \smallint \nolimits \mathop \prod \limits_{i = 1}^K d\textbf{r}_i^1d\textbf{r}_i^2O\left( {\left\{ {\textbf{r}_i^1,\textbf{r}_i^2} \right\}} \right)}}{{{V^{2K}}}}\]
To use Ewald's method, one should mimic the infinite space by sampling the system to the so-called "super cell" of volume $ L^3 $ with finite dyon density. Each super cell is represented 
by $ \textbf{n} \in {\textbf{Z}^3} $ , therefore $ \Phi (\textbf{r}) $ can be written as
\[\Phi (\textbf{r}) = \sum\limits_{\textbf{n} \in {\textbf{Z}^3}} {\sum\limits_j {\frac{{{q_j}}}{{|\textbf{r} - {\textbf{r}_j} - nL|}}} } ,\]
where $ j=(i,m) $. Then, one should split the above term to a "long-range part" and a "short-range part". The short-range part vanishes and the Fourier transform of the long-range part 
converges at spatial infinity.
Back to our work, we have calorons instead of dyons in the limit of $ \rho\gg\beta $ . The vector potential of the caloron in this limit is
\[{A_\mu }(\textbf{r}) =  - \frac{1}{2}{\sigma _3}\bar \eta _{\mu \nu }^3{\partial _\nu }\ln \Psi \left(\textbf{r}  \right),\]
where $ \Psi (\textbf{r}) = \frac{{\left| {\textbf{r} - {\textbf{z}_2}} \right| + \left( {\textbf{r} - {\textbf{z}_2}} \right).{\textbf{e}_3}}}{{\left| {\textbf{r} - {\textbf{z}_1}} \right| + \left( {\textbf{r} - {\textbf{z}_2}} \right).{\textbf{e}_3}}} $ and $ z_1 $ and $ z_2 $ represent the positions of two constituent 
monopoles. If the observing point is far from the caloron, then $ r \gg {z_1},{z_2} $. In the first approximation, where $ \frac{{{z^2}_{i}}}{{{r^2}}} $ and $ \frac{{{z^2}_{i}}}{r} $ ($ i=1,2 $) are negligible, we can write the vector potential as
\[{A_0}(\textbf{r}) = \left(\delta _{\mu 0}2\pi \omega T- \frac{1}{2}\sum\limits_{\textbf{n}\in {\textbf{Z}^3}} {\sum\limits_j {\frac{1}{{|\textbf{r} - {\textbf{r}_j} -\textbf{n} L|}}} }\right){\sigma _3},\]
Since we have a long-range vector potential, we can use the Ewald's method.
\\
Now we can consider the two interacting calorons as one object and write its vector potential as in eq.\ref{rotation} . This vector potential is also long-range and we can use Ewald's method for an ensemble of K non-interacting two-caloron objects with appropriate separation and repeat the procedure of this section two compute the correlation function of Polyakov loops.

\section{conclusion}
 \label{7}
In this research we try to suggest computing the potential between interacting calorons with non-trivial holonomies.The caloron-Dirac string interaction may be used to obtain the 
interacting potential between two calorons. Having this potential in hand, we have two ways to compute the correlation function and then the heavy quark potential. One way is calculating 
the partition function by computing the metric of the moduli space of the caloron ensemble and the other one is using the Ewald's method to include the effect of the long-range vector 
potential of calorons in the limit where $ \rho\gg\beta $ .

\begin{theacknowledgments}
We would like to thank M. Muller-Preussker and V. Petrov for the very useful discussions. We are grateful to the research council of the University of Tehran for
supporting this study.
\end{theacknowledgments}

\bibliographystyle{aipproc}

\end{document}